\documentstyle[preprint,aps,overcite]{revtex}
 
\begin{document}
\input psfig.tex
\title{
Regarding the Deceleration of the Universe
}
\author{Paul J. Steinhardt}
\address{
Department of Physics and Astronomy\\
University of Pennsylvania\\
Philadelphia, Pennsylvania 19104  USA
\\ Draft: \today
}

\maketitle

 In the standard big bang model, the expansion rate 
of the  Universe
is predicted to slow down as the Universe evolves.\cite{Wein}$^,$\cite{KT}
  The deceleration parameter, $q_0$,   measures 
the present deceleration rate.  
The deceleration  causes a   deviation from the linear relation between
distance and red shift (the Hubble law) at large red shifts.
A proposed approach for determining $q_0$ is to measure
the red shifts and distances of
Type Ia supernovae,\cite{super1}$^,$\cite{super2} 
whose luminosities may be calibrated 
so that the inverse-square-law can be used
 to judge their distance.
To date, the strategy  has   been  to use measurements at 
low red shift $z\le 0.1$ to precisely determine the linear distance-red 
shift relation\cite{Press}  and then focus
on measuring the deviation from the linear law\cite{super1}$^,$\cite{super2}
  using supernovae at high
red shifts
 $z= 0.35-0.6$ to determine $q_0$.

  This note makes a simple point that is important in interpreting the 
supernovae results: 
At red shift $z=0.35-0.6$,  the deviation from the linear Hubble
relation does not depend on $q_0$ alone, but also  on the matter-energy 
content  of the universe.  It is well-known that this dependence
on matter-energy content is significant at sufficiently high red shifts, 
but it has not been generally
 appreciated that the dependence is non-negligible
at red shifts as modest as $z=0.35$.

The effect is easy to understand.  Observations at non-zero red shift
measure properties of the universe at earlier times when the light
was emitted.  In some models, such as open models with low matter 
density, the deceleration rate at $z\approx 0.5$ is comparable to the 
deceleration rate today ($q_0$). In other models, such as those 
with non-zero cosmological constant, the universe is undergoing 
a fairly rapid transition from  a decelerating, matter-dominated universe at
$z=0.35$ towards an accelerated expansion in a  universe dominated by a 
non-zero cosmological constant. The two models predict a different 
deviation from the linear Hubble law at $z=0.35$ even if the present
deceleration rate $q_0$  is the same.

The ``luminosity distance" between a  given source and us is defined as
$d_L^2 \equiv  {\cal L}/ 4   \pi {\cal F}$
where ${\cal L}$ is the emitted energy per unit
time and ${\cal F}$ is the 
 energy received
per unit time.  
An elementary calculation\cite{Wein}$^,$\cite{KT} shows that 
$d_L=(1+z)r_1$ where the comoving distance $r_1$ satisfies:
\begin{equation} \label{eq4}
\int_0^{r_1} \frac{dr}{(1-kr)^{1/2}}= 
\int_0^z  \,\frac{dz' \, [H_0(1+z')]^{-1}}{ \left[ \Omega_m (1+z') + \Omega_{\Lambda}(1+z')^{-2} + \Omega_{\alpha} (1+z)^{1+3 \alpha} +
(1-\Omega_m-\Omega_{\Lambda}-\Omega_{\alpha})\right]^{-1/2}};
\end{equation}
here $H_0$ is the Hubble constant, and $\Omega_m$, $\Omega_{\Lambda}$ and $\Omega_{\alpha}$ are the 
ratios of the matter density, the cosmological constant,
and other possible forms of matter-energy, respectively, to the 
critical density. We have generalized to include 
possible  matter-energy with  general 
 equation-of-state $\alpha\equiv p/\rho$, where
$p$ is the pressure and $\rho$ is the energy density.
Then, $q_0=\frac{1}{2}\Omega_m -\Omega_{\Lambda} + 
\left(\frac{1+3 \alpha}{2} \right) \Omega_{\alpha}$.

Figures 1 and 2 illustrate the point. 
Figure 1 shows the distance-red shift relation for three different
models.
Note that  the  curves diverge from one another before $z= 0.35$ 
even though they correspond to identical $H_0$ and $q_0$.
Figure 2 is a blow-up of the range $z=0.35-0.6$  expressed in terms of
apparent magnitude ($m= 5 \, {\rm log}[d_L]+\, {\rm constant}$)  vs. red shift.
The Figure  magnifies the differences among the models in Figure 1, and shows
a model with $q_0=-0.25$ and $\Omega_{\Lambda}=\Omega_m=0.5$ that is nearly
indistinguishable from an open model with $q_0 \approx 0$.

Although the current strategy does not measure $q_0$ alone,
it provides a useful constraint on a combination of  $\Omega_m$, 
$\Omega_{\Lambda} $, $\Omega_{\alpha}$ 
and the spatial curvature.  If one imagines a multi-dimensional parameter-space with axes corresponding to each of these variables, a 
precise measurement at any given $z$ constrains models to some swath which 
depends on $z$.  Goobar and Perlmutter\cite{super3} have illustrated this for models with $\Omega_{\alpha}=0 $, and it is straightforward to generalize to other models.  
By precise measurements at disparate values of $z$, especially in the range $0.2$ to $0.5$, a set of constraints can be obtained which may make it possible to 
discriminate $q_0$ and the other parameters independently.

This work was supported in part by the US Department of Energy.  The author
wishes to thank S. Perlmutter,  M. Rees, J. Ostriker, W. Press, J. Bahcall, 
and E. Turner  for useful discussions.

\newpage
 
\begin{figure}
\centerline{\psfig{file=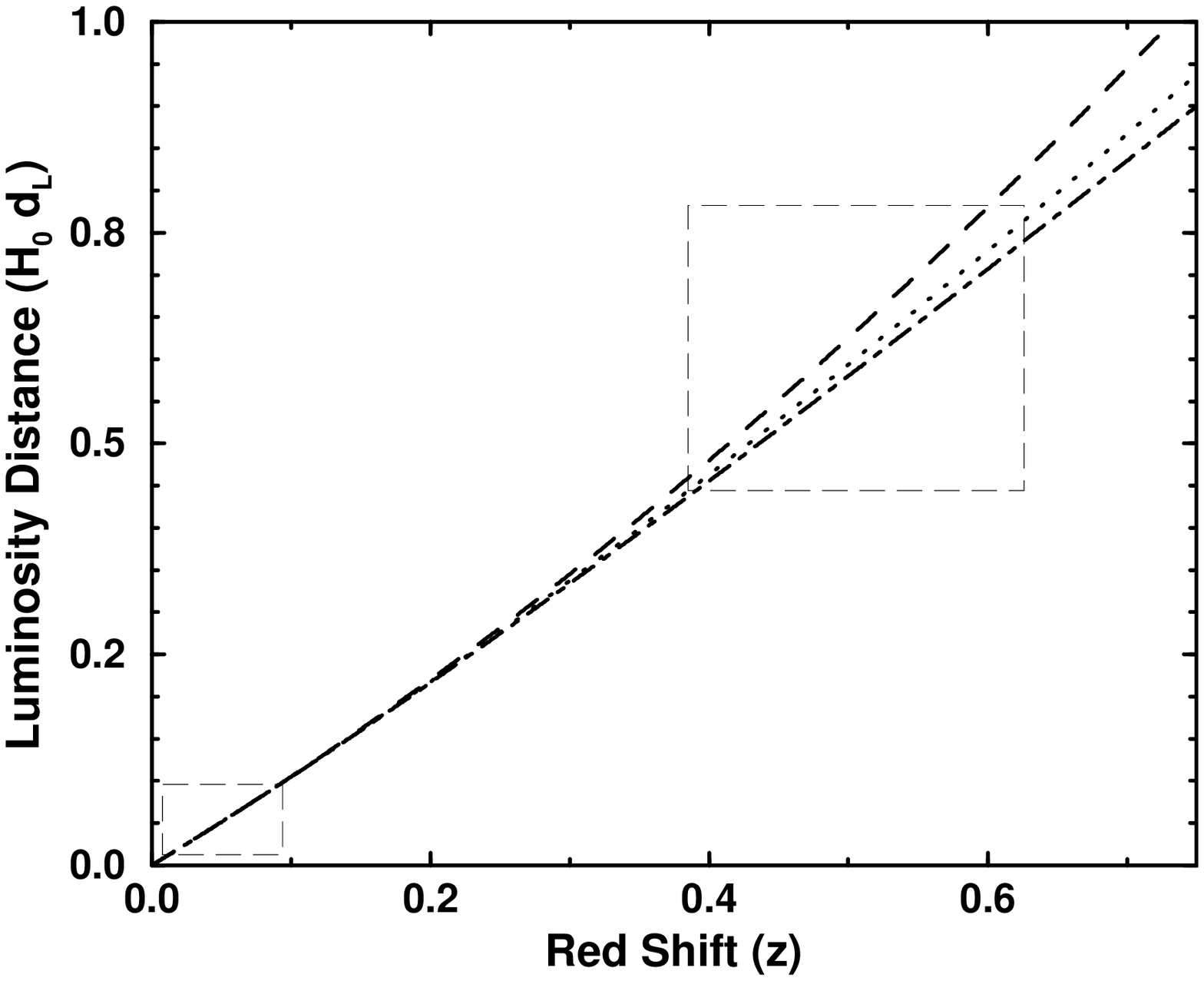,width=6.0in}}
\caption{ A demonstration that  the
luminosity distance vs. red shift curves can differ significantly
at $z \approx 0.5$ (upper
box)  for
models with same 
$q_0=0$: (a)  matter-dominated model (dashed) with $\Omega_m \rightarrow 0$,
$\Omega_{\Lambda}=\Omega_{\alpha}= 0$;  (b)  $\Omega_m=2/3$,
$\Omega_{\Lambda}=1/3$ and $\Omega_{\alpha}= 0$ (dotted); and (c) $\Omega_m=0.2$;
$\Omega_{\Lambda}=0.45$, and $\Omega_{\alpha}=.35$ (dot-dashed) 
with $\alpha=1/3$ 
representing hot, relativistic matter. 
}
\end{figure}

\begin{figure}
\centerline{\psfig{file=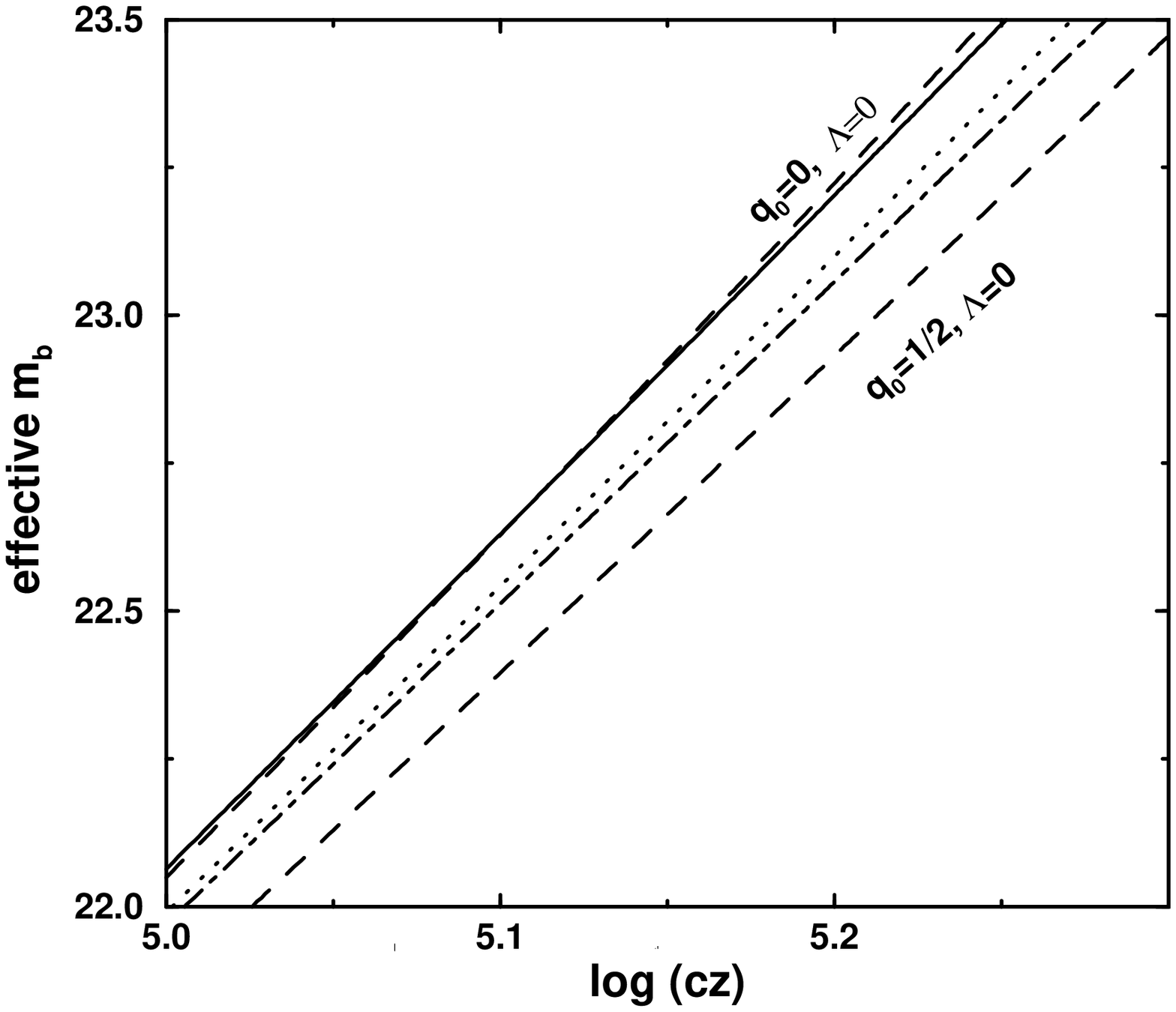,width=6.0in}}
\caption{
The red shift regime  
 $z=0.35-0.6$ replotted in terms of 
effective apparent magnitude.
The upper and lower dashed curves correspond to  $q_0=0$ and $q_0=1/2$
assuming $\Omega_{\Lambda}=\Omega_{\alpha}=0$.  The  dotted and 
dot-dashed curves (same as Figure 1) also have $q_0=0$, but differ 
significantly from the upper dashed curve.  The solid line corresponds
to $\Omega_{\Lambda}=\Omega_m=0.5$ and $q_0=-0.25$, yet it is 
difficult to distinguish from the upper dashed curve with $q_0=0$.
}
\end{figure}


\begin{thebibliography}{1234546}
\bibitem{Wein} Weinberg, S., {\it Gravitation and Cosmology}, (Wiley, New York,
1972), pp. 441-451.
\bibitem{KT} Kolb, E.W. and Turner, M.S., {\it The Early Universe},
(Addison-Wesley, Redwood City, 1990), pp. 41-45, 82-85.
\bibitem{super1} 
Perlmutter, S. {\it et al.}, {\it Ap. J.} 440, L41-44 (1995).
\bibitem{super2}
Perlmutter, S.  {\it et al}, astro-ph/9602122 (1996).
\bibitem{Press}
 Riess, A. G., Press, W. H., \& Kirshner, R. P. {\it Ap. J.} 438,
L17-20 (1995).
\bibitem{super3}
Goobar, A. and Perlmutter, S., {\it Ap. J.} 450, 15 (1995).
\end{thebibliography}
\end{document}